\documentstyle[prl,aps]{revtex}

\begin{document}

\draft
\title{The Speed of Gravity Has Not Been Measured From Time Delays}
\author{Joshua A.\ Faber}
\address{Department of Physics and Astronomy, 
 Northwestern University, Evanston, IL 60208}

\date{\today}

\maketitle

\begin{abstract}
The recent passage of Jupiter by the quasar QSO J0842+1835 
at a separation of 3.7 arcminutes
on September 8, 2002, combined with recent advances in interferometric radio
timing, has allowed for the first measurement of higher-order post-Newtonian 
terms in the Shapiro time delay which depend linearly on the velocity of the 
gravitating body.  Claims have been made that these measurements also allow
for the measurement of the propagation speed of the gravitational
force.  This conclusion disagrees with recent calculations done
in the parameterized post-Newtonian (PPN) model, which find no
dependence of the velocity-dependent terms in the time delay 
on the ``speed of gravity'' to the stated order.  
Here, to test out these claims and
counterclaims, we calculate the time delay in the limit of an
instantaneous gravitational force, 
and find that the velocity-dependent terms are in complete agreement with 
previous PPN calculations, 
with no dependence on the speed of gravity.  We conclude that
the speed of gravity cannot be determined by measuring these 
terms in the Shapiro time delay, and suggest a reason why other groups
mistakenly came to the opposite conclusion. 
\end{abstract}

\section{Historical Background}

Almost 40 years ago, I.I.~Shapiro pointed out that the time delay
which results from light appearing to slow down as it passes through a
gravitational potential could be measured within our solar system
\cite{ref:Shap1}.  The measurement was done three years later, and
provided a new test of the theory of general relativity (GR) \cite{ref:Shap2}.
Recent advances have now allowed radio astronomers to measure higher
order post-Newtonian (PN) terms in the Shapiro time delay, using
Very-Long Baseline Interferometry (VLBI).  A recent
passage of Jupiter by the position of a bright quasar at a separation
of only 3.7 arcminutes has allowed for some of these high-order terms
to be measured by Fomalont and Kopeikin, 
specifically those that have to do with the transverse
motion of Jupiter perpendicular to the line of sight to the
quasar \cite{ref:Fom}.  In a series of papers, Kopeikin has argued
that the terms in the Shapiro time delay which depend upon the velocity of
the gravitating body also depend upon the propagation speed of
gravity \cite{ref:Kop1,ref:Kop2}. 
Specifically, in Ref.~\cite{ref:Kop1}, he derives
an equation for the time delay is the form 
\begin{equation}  
t_d=-2\sum_a \frac{Gm_a}{c^3}\left(1-\frac{\vec{K}\cdot\vec{v}_a}{c_g}\right)
\ln (r_a-\vec{K}\cdot\vec{r}_a)+C, \label{eq:koptd}
\end{equation}
where $t_d$ is the time delay, $m_a$ and $v_a$ the mass and velocity
of the $a$'th gravitating body, respectively, $c_g$ is the propagation
speed of gravity, $\vec{r}_a$ the separation vector between the
gravitating body and the observer, $r_a$ its magnitude, 
C is an integration constant, and the vector $\vec{K}$ (called $\vec{N}$ by
Kopeikin) is given by
\begin{equation}
\vec{K}=\vec{k}-\frac{1}{c_g}\vec{k}\times(\vec{v}\times\vec{k}),
\label{eq:kopk}
\end{equation}
where $\vec{k}$ is a unit vector pointing in the direction of the
photon's path.  In what follows, for clarity, we will refer to
equations in Ref.~\cite{ref:Kop1} with the author's initial, thus our
Eqs.~\ref{eq:koptd} and \ref{eq:kopk} correspond exactly to Eqs.~(K22) and
(K23).  

Recently, Asada \cite{ref:Asada} has taken issue with the
interpretation of the velocity-dependent terms in the time delay for
models which assume that $c_g=c$.  Will goes further,
evaluating the time delay for models in which $c_g \ne c$, and concludes 
that there is an error in Kopeikin's
derivation of the time delay \cite{ref:Will}.  Using the parameterized
post-Newtonian (PPN) framework (see Ref.~\cite{ref:Will93} for more
details), he concludes that the proper time delay is given by 
\begin{equation}  
t_d=-2\sum_a \frac{Gm_a}{c^3}\left(1-\frac{\vec{K}\cdot\vec{v}_a}{c}\right)
\ln (r_a-\vec{K}\cdot\vec{r}_a)+C, \label{eq:willtd}
\end{equation}
where
\begin{equation}
\vec{K}=\vec{k}-\frac{1}{c}\vec{k}\times(\vec{v}\times\vec{k}).
\label{eq:willk}
\end{equation}
Using his initials for clarity as well, our Eqs.~\ref{eq:willtd} and
\ref{eq:willk} correspond exactly with Eqs.~(W35) and (W36), where we
have set several of the PPN parameters which do not affect the final
result equal to the values predicted by GR.

In what follows, we will repeat the calculations of Kopeikin and Will,
but in a model where the gravitational force is instantaneous,
i.e. $c_g\rightarrow \infty$.
We identify where the two methods agree under this assumption, 
and where they find different
mathematical expressions.  In short, we find that Will's result is
correct given his assumptions regarding the PPN framework.  Kopeikin's
result is inconsistent with his assumptions, revealing that his
derivation contains a mathematical error.  We identify the likely
cause of it in the course of our derivation.

\section{Calculating the Time Delay}

For the following calculations, we make use of the following
assumptions.  A photon travels along the x-axis of our coordinate system.  
It passes a body of mass $m_a$ moving with velocity
$\vec{v}_a=(v_{\parallel}, v_{\perp}, 0)$ along a straight line.  
The origin of the time coordinate is defined
such that the instantaneous 
separation vector between the gravitating body and the photon is
perpendicular to the photon's path at $t=0$.  This does not correspond
exactly to the moment of closest approach between the photon and the
gravitating body, which can easily be calculated to be first-order in
$v_a/c$.
Using these velocities, we find that the position of the 
of the gravitating body is given by $\vec{x}_a(t)=(v_{\parallel} t,
v_{\perp}t+y_0, z_0)$  and
that of the photon by $\vec{x}_p(t)=(ct, 0, 0)$.
Our observer is placed far from the
gravitating body at some distance $x_o$ along the photon's path, at
position $\vec{x}_o=(x_o, 0, 0)$.  As these calculations are
traditionally done by placing an observer at the barycenter of the
system, we should technically include at least one other gravitating
body, but we see immediately that it will have no effect to lowest
order on the measured time delay.

To calculate the time delay for the photon passage, we note that since
we can make the gravitating mass sufficiently small, we can ignore all
effects associated with the deflected path of the photon, which are of
higher order in $v_a/c$.  The equations of motion for
the photon's path as a function of its wave vector 
$k^{\mu}=dx^{\mu}/dt=(1,k^i)$
are completely determined from a description of the
spacetime metric, and the constraint equations for a null geodesic.
These constraint equations are given by both Eq.~(W11) and Eq.~(K14),
in complete agreement, in the form
\begin{eqnarray}
\frac{d^2 x_p^i}{dt^2} + k^{\mu}k^{\nu}(\Gamma^i_{\mu\nu}-k^i
\Gamma^0_{\mu\nu})&=&0 \label{eq:prop}\\
\left(\frac{ds}{dt}\right)^2=g_{00} +2g_{0i} k^i +g_{ij}k^i k^j
&=&0. \label{eq:normal} 
\end{eqnarray}

While Kopeikin and Will agree on the geodesic equation for the photon,
they disagree on the form of the metric.  Will, using the
standard parameterized post-Newtonian (PPN) formalism, assumes that
the metric, Eq.~(W9), takes the form 
\begin{eqnarray} 
g_{00} &=& -1+2U \label{eq:met0}\\
g_{0i} &=& -4 V_i \label{eq:metric}\\
g_{ij} &=& (1+2U)\delta_{ij}, \label{eq:met2}
\end{eqnarray}
where we have set the PPN parameters $\gamma$ and $\alpha_1$ equal to
the GR values $\gamma=1, \alpha_1=0$, and the retarded
Li\'{e}nard-Wiechert mass and momentum potentials are given by
Eq.~(W10) as 
\begin{eqnarray}
U(t, \vec{x}) &=& \frac{Gm_a}{c^2 r_e(t, \vec{x})} \label{eq:upot}\\
V_i(t, \vec{x}) &=& \frac{Gm_a \vec{v}_a}{c^3 r_e(t, \vec{x})},
\label{eq:vpot} 
\end{eqnarray}
where we define $r_e$ to be the ``effective'' distance of the
retarded potential, such that
$r_e(t,\vec{x})=|\vec{x}-\vec{x}_a(t')|$, where the ``emission''
time of the gravitational force must satisfy $t-t' = r_e/c_g$,
where $c_g$ is the speed of gravity.  

Kopeikin's metric is slightly different, in ways that deserve some
clarification.  He introduces a parameter $\tau$, which is used to
describe all quantities involving gravitation in his framework.  It is
defined such that $c_g \tau = ct$.  Unfortunately, the gravitational
equations treat this parameter as a physical time for gravitation.
Thus, the wave equation for linear metric perturbations, Eq.~(K5), reads
\begin{equation}
\left(-\frac{1}{c_g}^2\frac{\partial^2}{\partial\tau^2}+\nabla^2\right)
\bar{\gamma}^{\mu\nu}(\tau,\vec{x})=-\frac{16\pi G}{c^4}
\Theta^{\mu\nu} (\tau,\vec{x}),
\end{equation}
where $\bar{\gamma}^{\mu\nu}$ is the standard trace-reversed linear metric
perturbation, and $\Theta^{\mu\nu}$ is his modified stress-energy
tensor.  We note that for this equation to give a traveling wave
solution moving at speed $c_g$, we must interpret $\tau$ as the
physical time, not $t$.  We believe it is confusion between these two
quantities which leads to errors in his final conclusions.  In any
case, Kopeikin modifies not only the stress-energy tensor but the
metric perturbation as well, yielding a metric that agrees with Will's
(if we view $\tau$ as the physical time-variable) but for a different
momentum potential, given by Eq.~(K10) as
\begin{equation}
V_i(t, \vec{x}) = \frac{Gm_a (\vec{v}_a)_i}{c^2 c_g r_e(t, \vec{x})}.
\label{eq:vpot2}
\end{equation}

In the calculation that follows we will see that the differences in
the two metric formulations lead to different values for the
gravitomagnetic drag experienced by a photon passing by a moving
body.  We will also see that Will's calculation is completely
consistent, whereas Kopeikin's final answer is inconsistent with the
photon propagation equation.

We start our calculation by deriving the magnitude of the photon's
wavevector in our metric, Eq.~\ref{eq:metric}, noting that Kopeikin and
Will differ on the value of the momentum potential $V_i$.  We denote
the magnitudes of
the wave-vector $k^i=(k,0,0)$ and the momentum potential $V_i$ by $k$ and
$V$, respectively. The x-component of the momentum potential is
denoted $V_{\parallel}$.  From Eqs.~\ref{eq:normal}--\ref{eq:met2},
we find
\begin{equation}
(-1+2U) -8 V_{\parallel} k + (1+2U) k^2=0 \rightarrow
k\equiv
\frac{dx_p}{dt}=\frac{4V_{\parallel}+\sqrt{1-4U^2+16V_{\parallel}^2}}{1+2U}.
\end{equation}
Working only to 1.5PN order, we throw out all terms involving $U^2$ or
$V^2$, and find 
\begin{equation}
k=\frac{dx_p}{dt}=\frac{1+4V_{\parallel}}{1+2U}\sim (1-2U+4V_{\parallel}).
\end{equation}
To find the time delay, we integrate this expression over the path
length, finding for the time delay
\begin{equation}
t_d = (\int dt) - t_0 = \left[\int
\frac{dx_p}{c}\frac{1}{1-2U+4V_{\parallel}}\right] - 
c\delta x \sim  \int \frac{dx_p}{c} (2U-4V_{\parallel}),
\end{equation}
where $t_0$ is the time required to travel a distance $\delta x$ in
the absence of gravitating bodies.  The two terms here have entirely
different meanings, but both are easily understood.  The first,
$\propto \int U dx_p$, which
appears for static gravitational sources as well, is the simple
geometric time delay which results from photon's traveling through
any gravitational potential.  
The second term, $\propto \int V_{\parallel} dx_p$
is the gravitomagnetic contribution to the time delay.  In simplest
terms, it can be thought of as a ``frame-dragging'' term, whereby the
moving object pulls the photon along it's path.  At this point, it is
useful to compare our result with those of Will and Kopeikin.  To do
so, we make use of the ratio between the momentum potential and the
mass potential, which is constant throughout space and time in both
formalisms so long as the gravitating body moves at constant
velocity.  From Eqs.~\ref{eq:upot} and \ref{eq:vpot}, we see that in Will's
formalism that $V_i=U\slantfrac{(v_a)_i}{c}$, 
whereas from Eqs.~\ref{eq:upot} and
\ref{eq:vpot2}, Kopeikin's method yields ${V_i}=U\slantfrac{(v_a)_i}{c_g}$.
We find, respectively,
\begin{eqnarray}
t_d &\sim& 2 \left (1-\frac{2\vec{k}\cdot \vec{v}_a}{c}\right)\int
\frac{dx_p}{c} U~~~ {\rm (Will)} \label{eq:tdwill}
\\
&\sim& 2 \left (1-\frac{2\vec{k}\cdot \vec{v}_a}{c_g}\right)\int
\frac{dx_p}{c} U~~~{\rm (Kopeikin)}, \label{eq:tdkop}
\end{eqnarray}
which correspond with Eq.~(W16), and Eq.~(K21).

Calculating the time-delay integral is straightforward, so long as we calculate
the effective distance correctly for our chosen value of $c_g$.  To
simplify matters, and emphasize the difference between the frameworks
used by Will and Kopeikin, we will evaluate the time delay for a model
with instantaneous gravitational propagation, i.e., $c_g \rightarrow
\infty$.  In Kopeikin's notation, this represents the limit $\epsilon
\rightarrow 0$, which he refers to as the Newtonian limit.  Given the
trajectories of the photon and gravitating body stated above, we find
that the effective radius is given as a function of time as 
\begin{eqnarray*}
r_e(t)&=&\left[(c-v_{\parallel})^2 t^2 +
(v_{\perp}t+y_0)^2+z_0^2\right]^{0.5}\\ 
&=&\left[(c^2-2cv_{\parallel}+v_a^2)t^2+2y_0 v_{\perp} t+y_0^2+z_0^2\right]^{0.5}.
\end{eqnarray*}
To simplify the expression, we note that at the moment of closest
approach, at $t_{min}=-y_0 v_{\perp} t/[(c^2-2cv_{\parallel}+v_a^2)t^2]$,
the distance reaches its minimum value of $r_{min}\equiv
r_e(t_{min})$.
At all other times, the effective distance is given by
\begin{equation}
r_e(t)=\sqrt{r_{min}^2+|(\vec{u}c(t-t_{min}))|^2},
\end{equation}
where $\vec{u}=\vec{k}-\vec{v}_a/c$, $\vec{k}$ is the unit 3-vector pointing
along the photon's path, and we know that $\vec{r}_{min}\cdot\vec{u}=0$.
Note that the appearance of $c$ in these expressions has nothing to do
with assumptions about the speed of gravity, it merely represents the
speed of the photon when we calculate relative distances and velocities.

The calculation of the time delay will allow for a direct test of Will
and Kopeikin's results, since both derive the same integral for the
time delay, albeit with different prefactors, as we found in
Eqs.~\ref{eq:tdwill} and \ref{eq:tdkop}.
The time delay integral is analytic, since we can integrate along the
unperturbed path of the photon.  Changing variables to $dt=dx_p/c$,
shifting the time axis to a new time variable offset by $t_{min}$,
and defining $u=|\vec{u}|$, we find
\begin{eqnarray}
t_d&\sim&\int\frac{dt}{U(t)} \sim Gm_a \int \frac{dt}{r_e(t)} \sim
Gm_a \int \frac{dt}{\sqrt{r_{min}^2+u^2c^2t^2}} \\
&\sim& Gm_a\left(\frac{1}{uc}\right)\ln[\sqrt{r_{min}^2+u^2c^2t_f^2}-uct_f],
\label{eq:tdint}
\end{eqnarray}
where $t_f$ is the arrival time of the photon.  Note that in deriving
this expression, we made no assumptions about the geometry of the
system at the moment of closest approach, which Kopeikin discusses in
his appendix. In what follows, we will
only use the fact that vectorial velocity difference between the
photon and the gravitating body is perpendicular to the separation
vector at closest approach.

We start with an analysis of the prefactor $1/uc$.  Calculating this
to first order in $\vec{w}_a\equiv \vec{v}_a/c$ shows us that
\begin{eqnarray}
\frac{1}{uc}&=&\frac{1}{c|\vec{k}-\vec{w}_a|}=
\frac{1}{c\sqrt{\vec{k}\cdot \vec{k}-2\vec{k}\cdot\vec{w}_a+
\vec{w_a}\cdot\vec{w}_a}}
=\frac{1}{c\sqrt{1-2w_{\parallel}+w^2}} \\
&\sim& \frac{1}{c(1-w_{\parallel})}
\sim\frac{1}{c}(1+w_{\parallel})=\frac{1}{c}(1+v_{\parallel}/c)
=\frac{1}{c}(1+\frac{\vec{k}\cdot\vec{v}_a}{c}).
\label{eq:uc}
\end{eqnarray}
This multiplicative factor is derived directly from the integral of
the gravitational potential along the line of sight, and has
absolutely nothing to do with the speed of gravity, which we are
assuming here to be instantaneous.  It is essentially a Doppler
correction, representing the increased (decreased) time that the photon
spends in the deepest part of the gravitational potential because of
the gravitating body's parallel (anti-parallel) velocity component.
Comparing to Eq.~\ref{eq:tdwill}, we see that the ``Doppler'' term is half
the magnitude and opposite in sign from the gravitomagnetic correction
term.  On the other hand, this result is at odds with the
gravitomagnetic term derived by Kopeikin in Eq.~\ref{eq:tdkop}, which
contains $c_g$ in the denominator.  We will discuss this error below.

Moving on to the logarithmic term in the integral, we 
define the separation vector $\vec{r}(t)\equiv
\vec{x}_p(t)-\vec{x}_a(t)=\vec{r}_{min}+ct\vec{u}$ whose length is
$r_e(t)=\sqrt{r_{min}^2+u^2c^2t^2}$.  Additionally, we note that
\begin{equation}
\vec{k}\cdot\vec{r}(t)=\vec{k}\cdot\vec{r}_{min}+
ct(\vec{k}\cdot\vec{k}-\vec{k}\cdot\vec{w_a})=
\vec{k}\cdot\vec{r}_{min}+ct(1-\frac{\vec{k}\cdot\vec{v_a}}{c}). \label{eq:krmin}
\end{equation}
This expression is similar to the term $uct$ which appears in 
logarithmic term of the time delay, but not equal to it at the
required order because of the appearance of a term
$\vec{k}\cdot\vec{r}_{min}$.  However, we see that 
\begin{equation}
\vec{r}_{min}\cdot\vec{u}=0 \rightarrow
\vec{r}_{min}\cdot\vec{k}-\vec{r}_{min}\cdot\vec{w_a}=0.
\end{equation}
Denoting by $r_{\parallel}$ and $r_{\perp}$ the components of
$\vec{r}_{min}$ parallel to $\vec{v}_{\parallel}$ and
$\vec{v}_{\perp}$, respectively, we find, to lowest order,
\begin{equation}
r_{\perp}w_{\perp}=\vec{r}_{\parallel}(1-w_{\parallel})\sim
\vec{r}_{\parallel}=\vec{r}_{min}\cdot\vec{k}.
\end{equation} 
Thus, we can subtract off the term $\vec{r}_{min}\cdot\vec{k}$ in
Eq.~\ref{eq:krmin} by taking the product of the transverse velocity of
the gravitating body with the minimum separation vector.  It is
trivial to deduce that
$\vec{k}\times(\vec{v}_a\times\vec{k})=\vec{v}_{\perp}$.  We see
though, that to first order in $v_a/c$,
\begin{eqnarray}
\vec{w}_{\perp}\cdot\vec{r}(t)&=&\vec{w}_{\perp}\cdot\vec{r}_{min}+
ct(\vec{w}_{\perp}\cdot\vec{k}-\vec{w}_{\perp}\cdot\vec{w}_a) \\
&\sim&w_{\perp}r_{\perp},
\end{eqnarray}
where the second term vanishes from orthogonality, and the last
because it is second-order in $v_a/c$.  Combining these results, we
conclude
\begin{eqnarray}
\ln[\sqrt{r_{min}^2+u^2c^2t_f^2}-uct_f]&\sim&
\vec{r}(t)-ct\left(1-\frac{\vec{k}\cdot\vec{v}}{c}\right) \\ 
&\sim& r(t)-\left(\vec{k}-\frac{1}{c}\vec{k}\times(\vec{v}_a
\times\vec{k})\right)\cdot \vec{r}(t). 
\label{eq:final}
\end{eqnarray}

Plugging Eqs.~\ref{eq:tdint}, \ref{eq:uc}, and \ref{eq:final} into
Eq.~\ref{eq:tdwill} we recover Eqs.~(W35) and (W36).  Plugging the
same three equations into 
Eq.~\ref{eq:tdkop} does not recover Kopeikin's Eqs.~(K22) and (K23), which
have the speed of gravity $c_g$ where we have the speed of light $c$.  We
conclude that his calculation is erroneous, since the time-delay
integral we calculated had nothing to do with the speed of gravity,
which we took to be instantaneous.  It is also independent of all
metric quantities other than the gravitational potential which appears
in $g_{00}$.  Why then does he find different numbers than ours?

A clue is provided by the fact that our results (as well as those of
Will, whose results ours reproduce) can be brought into agreement with
Kopeikin's if we replace $v_a$, the velocity of the gravitating body,
with $v_a c/c_g$.  In his notation, this is the transformation $v_a\rightarrow
\epsilon v_a$.  We believe the error he makes is to use $\tau=\epsilon
t$ as the physical time for the gravitating object, declaring that its
physical velocity is given by $d\vec{x}_a/d\tau=\vec{v}_a$, while
keeping $t$ as the time variable used to describe the photon's motion,
as in Eq.~(K13).  If we convert the velocity he uses for the
gravitating object from $\tau$-based coordinates to t-based
coordinates, we find immediately that
$\vec{v}_a(t)=d\vec{x}_a/dt=d\vec{x}_a/d\tau(d\tau/dt)=\epsilon
\vec{v}_a$, which corresponds to the unphysical and erroneous velocity
found in his calculations.  

\section{Conclusions}

We have repeated the calculations performed by Will and Kopeikin, in a
model where the speed of gravity is instantaneous.  We find that 
first-order terms in the time delay resulting from the motions of
gravitating bodies are independent of the speed of gravity, and follow
the form written down by Will in every respect.  We find in addition
that Kopeikin's equation for the time delay, which asymptotically
approaches the time-delay for static bodies in the limit of an
infinite speed of gravity, is wrong.  We conclude that since the
first-order velocity terms in the time delay are independent of the
speed of gravity, Fomalont and Kopeikin's high-precision measurements
of the time delay of light from quasar QSO J0842+1835 passing by the
edge of Jupiter on September 8, 2002 \cite{ref:Fom}, while a truly
impressive observational feat in radio astronomy, 
provided a measure of the speed of light only, not the speed of
gravity as was claimed.

The author wishes to thank P. Grandcl\'ement and E. Bertschinger for
helpful discussions.  He also wishes to thank S. Kopeikin for
clarifying  discussions about the details of his results. 

\end{document}